\documentclass[12pt,a4paper]{article}
\usepackage[pdfstartview=FitH,colorlinks=true,linkcolor=black,anchorcolor=black,citecolor=black,urlcolor=black]{hyperref}
\usepackage[english]{babel}
\usepackage{amsmath,amssymb,titling,authblk}
\usepackage{slashed}
\usepackage{amsmath}
\usepackage{amscd}
\usepackage[normalem]{ulem}
\usepackage{appendix}
\usepackage{bbold}
\usepackage{bm}

\usepackage{pdflscape}
\usepackage{yfonts}
\usepackage{amscd}

\usepackage{epsfig}
 \usepackage{microtype}
 
 \usepackage{hyperref}

\usepackage{cite}

\allowdisplaybreaks[4]

\usepackage {color}
\usepackage{bbold}
\usepackage{braket}
\usepackage{bbm}

\advance\hoffset by -1.1truecm
\advance\voffset by -1.0truecm
\newcommand{\be}{\begin{equation}}
\newcommand{\ee}{\end{equation}}
\newcommand{\bea}{\begin{eqnarray}}
\newcommand{\eea}{\end{eqnarray}}

\newcommand{\del}{\partial}
\newcommand{\dd}{\mathrm d}

\usepackage{mathrsfs}
\usepackage{authblk}

\usepackage{float}
\restylefloat{figure}
\newcounter{appendice}

\newcommand{\formula}[1]{\be #1 \ee}
\newcommand{\formu}[1]{$  #1$}

\newcommand{\Tr}[1]{\:{\rm Tr}\,#1}
\newcommand{\tr}[1]{\:{\rm tr}\,#1}

\begin{document}

\setlength{\droptitle}{-6pc}

\title{Points. Lack thereof}

\renewcommand\Affilfont{\itshape}
\setlength{\affilsep}{1.5em}

\author[1,2,3]{Fedele Lizzi\thanks{fedele.lizzi@na.infn.it}}
\affil[1]{Dipartimento di Fisica ``Ettore Pancini'', Universit\`{a} di Napoli {\sl Federico~II}, Napoli, Italy}
\affil[2]{INFN, Sezione di Napoli, Italy}
\affil[3]{Departament de F\'{\i}sica Qu\`antica i Astrof\'{\i}sica and Institut de C\'{\i}encies del Cosmos (ICCUB),
Universitat de Barcelona, Barcelona, Spain}

\date{\sl Proceedings of the XXII Krakow Methodological Conference\\ \textit{Emergence
of the Classical}\\Copernicus Centre, 11-12 October 2018}

\maketitle 

\begin{abstract}\noindent
I will discuss some aspects of the concept of ``point'' in quantum gravity, using mainly the tool of noncommutative geometry. I will argue that at Planck's distances the very concept of point may lose its meaning. I will then show how, using the spectral action and a high momenta expansion,  the connections between points, as probed by boson propagators, vanish. This discussion follows closely~\cite{Kuliva} (Kurkov-Lizzi-Vassilevich Phys.\ Lett.\ B {\bf 731} (2014) 311, 
  [arXiv:1312.2235 [hep-th]].
\end{abstract}
\newpage

In this contribution I will discuss, from the point of view of a physicist\footnote{I have provided an extensive literature, which is however by all means not complete. I often referred to my own work, since it represents better the point of view presented here. For other relevant papers one may consult the references of the cited papers.} , a very classical concept: that of a \emph{point}. Although the concept is pervasive in physics and mathematics, as it often happens with the concept we think we know, its most profound meaning (beyond definitions) is far from easy. We encounter points very early in the study of mathematics I remember that in my  elementary school book a point was defined as: \emph{A geometrical entity without dimension}. 
I must confess that, after reading this definition, I was none the wiser about what a point is. The main reason could have been that I  was convinced I knew what a point is. I could produce them at will with my biro. Or, better, with a sharper pin, or an even sharper object. In any case I could envisage a limiting process for which I could always find something more ``pointlike''.

Point are crucial in geometry, and Euclid himself gave a definition of them: \emph{"That which has no part"}. This is not always true, often we just find expedient to ignore possible internal structures and talk of points, indeed it may be that there are ``points'' which have a rich structure, which we ignore for the problem at hand. 
In astrophysics, for example, a point may be a galaxy, or even a cluster of galaxies. 
In general relativity the set of point is usually the set of possible localise \emph{events}. This certainly implies some structure. 
In classical ``point particle'' dynamics we use points of phase space to describe the state of motion of a particle, which we imagine pointlike. What we have in mind when we talk of points, or pointlike particles, is always the limiting process I was alluding before. There may be reasons, such as technological limitations or convenience, to consider pointlike something which is not, but at the end of the limiting process, operationally, at the bottom there are points.

It is well known that for particle phase space (the space of positions and momentum/velocity) this vision becomes untenable when one considers quantum mechanics. It is impossible to know at the same time position of momentum of a particle. This is the content of Heisenberg \emph{uncertainty principle}~\cite{Heisenberguncert}
\be
\Delta p \Delta q \geq \frac\hbar2 \label{Heisenberg}
\ee
For quantum spacetime we have a well formed, successful theory, which is supported by a large body of experimental evidence, which we call \emph{Quantum Mechanics}. Although it may be formulated in several ways, by far the most useful, and rigorous, one is to consider all observables, and in particular position and moments, to be part of the algebra of \emph{operators}. The notion of point, and with it that of trajectory, is not present anymore in the theory. Every manual of quantum mechanics at some stage attempts a connection with classical physics, see for example~\cite[Sect.~II.4]{Messiah}. But the notion of point of phase space is just  an ill defined quantity in quantum mechanics. The closest we may get to it is the possibility to have a \emph{coherent state}. Independently on the formal group theoretical definition (see for example~\cite{Perelomov}, for our purposes it suffices the property that they are \emph{maximally localised} states, which will saturated the Heisenberg uncertainty bound~\eqref{Heisenberg} with the equality. A central role is played by the presence of a \emph{dimensional} quantity (Planck's constant $\hbar$) which acts as cutoff on phase space, thus avoiding the ``ultraviolet catastrophe".

Phase space has become a \emph{noncommutative geometry}, still it is possible to study the geometrical properties of such spaces, and this work has been pioneered by Alain Connes~\cite{Connesbook}. The idea is to rewrite ordinary geometry in algebraic terms, for example substituting the category of topological spaces with that of $C^*$-algebras. A physicists would say that we probe spaces via the fields built on it. For commutative spaces points are pure states, i.e.\ normalised linear maps from the algebra to complex, which have to satisfy certain properties. Once everything is rendered in an algebraic way, it is then possible to generalise to the noncommutative settings. The points of classical phase space, described by the commutative algebra generated by $q$ and $p$, were described by pure states, in the quantum setting the algebra is noncommutative, and the pure states are the wave functions. 

In quantum mechanics however configuration space remains ``classical'', and if one is willing to renounce to the information about momentum, positions remain the same ad in classical mechanics. 
I will not dwell further on quantum phase space, in the rest of this talk I will be concerned with ordinary (configuration) space, and spacetime. The possibility to consider quantum properties of spacetime goes back to Heisenberg himself, who was worried about the infinities of quantum field theory, seen as a new ultraviolet catastrophe, which at the time were considered a big problem. He wrote this in a letter to Peirleis, the latter told it to Pauli (who mentions it in his correspondence~\cite{HeisenbergtoPeirleis}). Some years afterwards, independently, Bronstein in 1938~\cite{Bronstein} noticed that in a theory containing both quantum mechanics and gravity, the presence of a quantity with the dimension of a length, Planck's length, would create problems, in principle not very different from the ones of quantum mechanics. If we include gravity in the game things change. We now have a length scale obtained combining the speed of light, Planck's constant and Newton's constant:  
\be
\ell=\sqrt{\frac{\hbar G}{c^3}}\simeq 10^{-33} \mathrm{cm}. \label{ell}
\ee
There was no follow-up of these ideas at the time, probably also because Bronstein was arrested not much after writing the paper by Stalin's police, and executed immediately. The phenomenon was presented more recently, and independently, in a very terse way by Doplicher, Fredenhagen and Roberts in 1994~\cite{DFR}.

I will present a caricature of these arguments, which hopefully captures the main idea in a nontechnical way.
It is a variant of the Heisenberg microscope justification of the uncertainty principle. The former goes as follows:in order  to ``see'' something small, of size of the order of \formu{\Delta x}, we have to send a ``small'' photon,
that is a photon with a small wavelength \formu{\lambda}, but a
small wavelength means a large momentum \formu{p=h/\lambda}. In
the collision there will a transfer of momentum, so that we can
capture the photon. The amount of momentum transferred is
uncertain. The calculations can be done in a more formal way, using the resolving power of an ideal  microscope to:
\be
\Delta p \Delta q \geq h
\ee
where $h=2\pi\hbar$ is the original Planck's constant.  The argument is very heuristic, and the result is indeed off by an order of magnitude (\formu{4\pi}). We know that in order to obtain the uncertainty principle it is necessary to have a solid theory, quantum mechanics, where $p$ and $q$ become operators, and then it is possible to prove rigorously~\eqref{Heisenberg}.

We are interested only is space, and not momentum, for which there is no limitation in quantum mechanics to an arbitrary precise measurement of $x$. We also change our notation to remark the difference with the previous discussion.
 In order to ``measure'' the position of an object, and hence the
``point'' in space, one has use a very small probe, which has to be very energetic, but on the other
side general relativity tells us that if too much energy is
concentrated in a small region a black hole is formed. In~\cite{DFR} the following relations were obtained
\bea
\Delta x_0(\Delta x_1+\Delta x_2+\Delta x_3)&\geq&\ell^2 \nonumber\\
\Delta x_1\Delta x_2+\Delta x_2\Delta x_3+\Delta x_1\Delta x_3&\geq&\ell^2
\eea
For a rigorous statement we would need a full theory of {\bf quantum gravity}. A theory which do not (hopefully yet) posses.

For geometry we need more than points, we need to know how to relate them, we need topology, metric, correlation among fields \ldots Several mathematical results of Gelfand and Naimark establish a duality between (ordinary) Hausdorff topological spaces, and $C^*$-algebras (for a quick review see~\cite[Chapt.~6]{WSS} and references therein). The $C^*$-algebra provides not only a set of points, but  that one may also infer topology, i.e.\ when a sequence of points converges to another point. If commutative algebras describe ordinary spaces, then \emph{noncommutative algebras} will describe ``pointless'' noncommutative spaces. This is the foundational principle of \emph{Noncommutative Geometry}. Even if we accept the idea that space is noncommutative, we must require that in analogy with phase space, ordinary space must be recovered when some parameter, $\ell$ in this case, goes to zero. This led to the introduction of \emph{deformed} algebra~\cite{Gerstenhaber} and $*$-products~\cite{starprod}. Of those the most famous one is the Gronewol-Moyal one~\cite{Gronewold,Moyal}, which was also introduced in string theory~\cite{FrohlichGawedski, LLS1, SeibergWitten}. 

In a deformed $*$-product, be it the original Gronewold-Moyal or one of its variations~\cite{GalluccioLizziVitale1, GalluccioLizziVitale2,TanasaVitale} it is still possible to define ordinary, point dependent functions, but the product is deformed. This has led to an interesting philosophical discussion as to the ``reality'' of points in such noncommutative geometries. We refer to the work of Huggett~\cite{Huggett} and its references, but I will move to considerations which involve the most advanced theory which encompasses relativity and quantum mechanics, \emph{quantum field theory}~\cite{Weinberg}, to infer what the relation among points are at very high energy. In particular I will use the knowledge form field theory at energies below the (yet to be defined) transition scale at which quantum geometry appears , to infer some knowledge of quantum spacetime. I envisage some sort of phase transition relating classical and quantum spaces, although this view is not necessary for the considerations I make below.
I will be very much inspired by noncommutative geometry, and I will be in a definite context, that of \emph{spectral geometry}, and especially the spectral action, but the reasoning I will make can be more general. A connection between spectral geometry and the $*$-product  is given by the fact that there is basis in which these products are represented by matrices~\cite{LizziVitaleReview}.

The way one can learn what happens beyond the scale of an experiment is to use the renormalization flow of the theory. We know that the coupling constants, i.e. the strength of the interaction, change with energy in a way which depends on the interacting fields and the (fermionic) particles present in the spectrum. This flow can be calculated perturbatively using data gathered at attainable energies, and then extrapolated at higher energies. The extrapolation will of course be valid only if no other, presently unknown, particles and interactions appear. Conventionally it is said that the calculation is valid ``in the absence of new physics''. Presently the known running of the three gauge interactions (strong, weak and electromagnetic) is presented in the figure. Gravitational interaction is not included since it does not give rise to a renormalizable interaction (and hence the need for quantum gravity!)
\begin{figure}[htbp]
\begin{center}
\includegraphics[scale=.45]{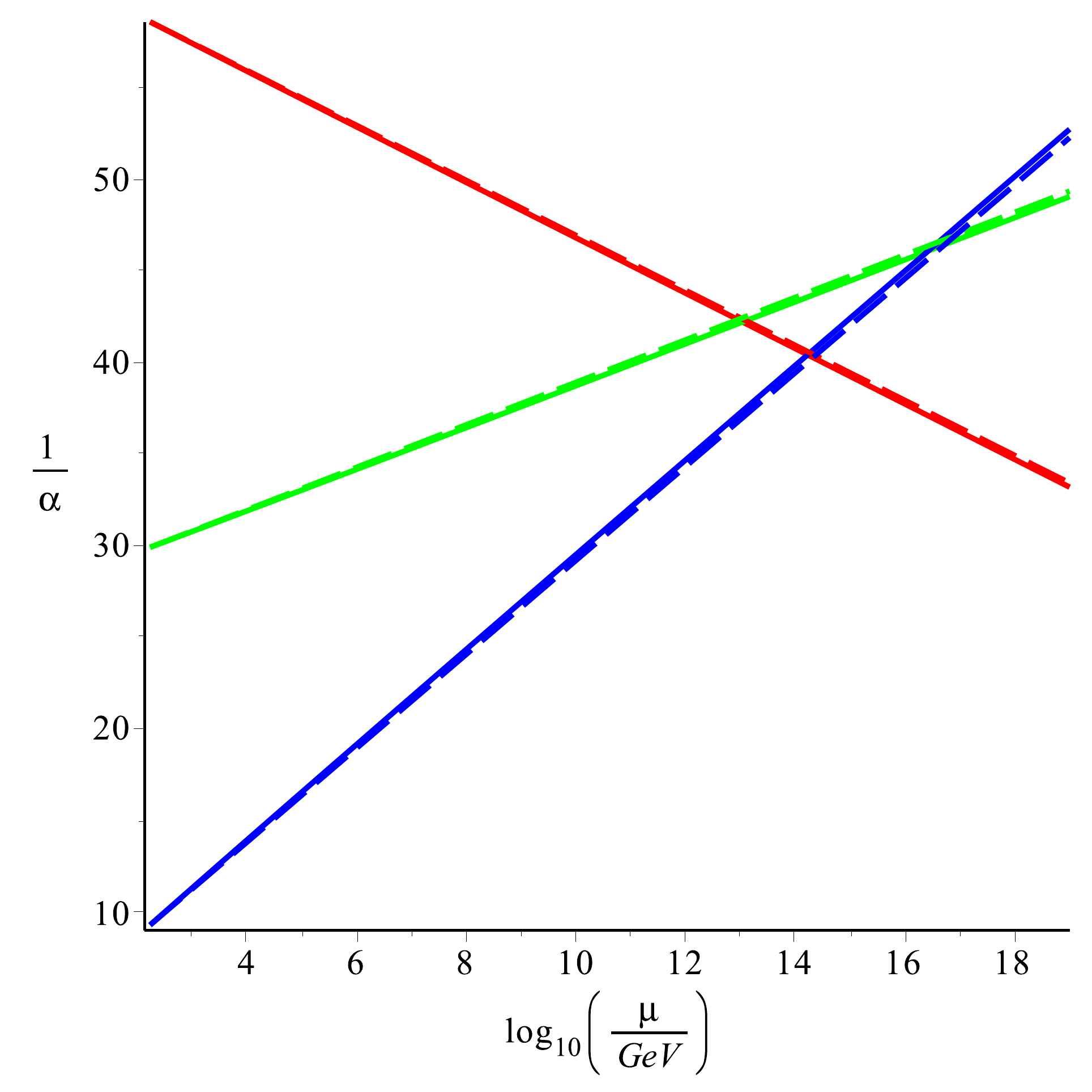}
\end{center}
\caption{\sl The running of the coupling constants of the three gauge interactions.}
\end{figure}
The boundary values at low energy are established experimentally, and the renormalization flow show that the nonabelian interactions proceed towards asymptotic freedom, while the abelian one climbs towards a Landau pole at incredibly high energies \formu{10^{53}~\mathrm{GeV}}. The figure also show that the three interaction \emph{almost} meet at a single value at a scale around $10^{15}$~GeV. The lack of a unification point was one of the reasons for the falling out of fashion of GUT's, but it should be noted that some supersymmetric theory allow the unification point. It is however unlikely that the quantum field theory survives all the way to infinite energy, or at least to the Landau pole. We all believe that this running will be stopped by {```something''} at the Planck energy (mass), which is the energy equivalent of the Planck's lenght $\ell$ of ~\eqref{ell}:
\be
M_p=\sqrt{\frac{\hbar c}c}\simeq
10^{19}~\mathrm{GeV}
\ee
 This unknown something we call \emph{quantum gravity}. At this scale there will certainly be some new physics, because it will be impossible to ignore gravitational effects.

I take the point of view that in the Planckian energy regime there is a fundamental change of the degrees of freedom of spacetime. Like what happens in a phase transition. One useful tool to describe this is new physics is \emph{Noncommutative Spectral Geometry}\footnote{For personal reviews in increasing level of detail see~\cite{Vilasiproc, Corfuproc, DevastatoLizziKurkov}.}. In this framework geometry is algebrized, the topological, metric and gauge aspects of the theory are encoded in a $C^*$-algebra acting on the Hilbert space of physical fermions, a generalised Dirac operator containing the information of the masses (Yukawa couplings) and of the metric, and chirality and charge conjugation. The model is successfully applied to the standard and it has some predictive power~\cite{AC2M2, ColdPlay, CCvS, Grandproc, Aydemir:2015nfa, Walterbook}.

The metric and geometric properties are encoded in the (generalized) Dirac operator $D$ which fixes the background around which expand the action. Here I must make a ``confession''. The model is Euclidean and it consider a compact space. The latter is usually considered a minor problem, but the infrared frontier may have surprises, see for example the recent interest in it in~\cite{Strominger, eomconstraints, scent}. The issue of a Lorentzian, or at least causal, version of this noncommutative geometry is under active investigation, a partial list of references is~\cite{Dungen:2015pca, Bizi:2016lbv, Franco:2012er, Franco:2013gxa, Franco:2015qra, DAndrea:2016hyl, Kurkov:2017wmx, Devastato:2017rlo, Bochniak:2018ucd, Aydemir:2019txw}

The eigenvalues of the Dirac operator on a curved spacetime are diffeomorphism-invariant functions
of the geometry. They form an infinite set of observables for general relativity and are therefore well suited to investigate the structure of spacetime. The interaction among fields is described by the \emph{Spectral Action:}
\be
S=\Tr\chi\left(\frac{D_A^2}{\Lambda^2}\right) \label{bosonicspectralact}
\ee
where $\chi$ is a cutoff function, which we may take to be a decreasing exponential or a smoothened version of the characteristic function of the interval, $\Lambda$ is a cutoff scale without which the trace would diverge. $D_A=D+A$ is a fluctuation of the Dirac operator, $A$ being a connection one-form built from $D$ as $A=\sum_i a_i[D,b_i]$ with $a_i,b_i$ elements of the algebra, the fluctuations are ultimately the variables, the fields of the action. the general ideas have a broader scope.  The presence of $\Lambda$ causes only a finite number of eigenvalues of $D$ to contribute, a finite number of modes. Finite mode regularization, based on the spectrum of the wave operator, was introduced in QCD~\cite{AndrianovBonora1, AndrianovBonora2, Fujikawabook}

The bosonic spectral actioncan be seen as a consequence of the spectral cutoff~\cite{Andrianov:2010nr,Andrianov:2011bc,Kurkov:2012dn}, for description of Weyl anomaly and also phenomena of induced Sakharov Gravity~\cite{Sakharov} and cosmological inflation\cite{Kurkov:2013gma}. It can also be seen as a zeta function calculated in zero~\cite{zeta}. The zeta-spectral action opens an intriguing opportunity to give a classically scale invariant formulation of the spectral action approach, where all the scales are generated dynamically.  Various mechanisms of scale generation were considered in both gravitational (Sakharov) or scalr fields sectors~\cite{ripples}. An enhanced role of the spectrum of the Dirac operator goes  far beyond the scope of spectral action. A simple analysis of the spectrum of the \emph{free} Dirac operator allows to arrive to a correct relation between three and four dimensional parity anomalies in gauge~\cite{Kurkov:2017cdz} and gravitational~\cite{Kurkov:2018pjw} sectors.

The spectral action can be expanded in powers of $\Lambda^{-1}$ using standard heat kernel techniques.
In this framework it is possible to describe the action of the standard model.
One has to choose as $D$ operator the tensor product of the usual Dirac operator on a curved background  $\slashed\nabla$ times a matrix containing the fermionic parameters of the standard model (Yukawa couplings and mixings), acting on the Hilbert space of fermions.In this way one ``saves'' one parameter, and can predict the mass of the Higgs. The original prediction was 170~GeV, which is not a bad result considering that the theory is basically based on pure mathematical requirements. When it was found at 125~GeV it was realized that the model had to be refined (right handed neutrinos play a central role) to make it compatible with present experiments. 
I will not dwell further on the Higgs issue, and concentrate on the role of the spectral action for spacetime.

Technically~\cite{manual}the bosonic spectral action is a sum of residues and can
be expanded in a power series in terms of $\Lambda^{-1}$ as
\be
S_B=\sum_n f_n\, a_n(D_A^2/\Lambda^2)
\ee
where the $f_n$ are the momenta of $\chi$
\begin{eqnarray}
f_0&=&\int_0^\infty \dd x\, x  \chi(x)\nonumber\\
f_2&=&\int_0^\infty \dd x\,   \chi(x)\nonumber\\
f_{2n+4}&=&(-1)^n \del^n_x \chi(x)\bigg|_{x=0} \ \ n\geq 0
\end{eqnarray}
the $a_n$ are the Seeley-de Witt coefficients which vanish for $n$
odd. For $D_A^2$ of the form
\be
D^2=-(g^{\mu\nu}\del_\mu\del_\nu\mathbb I+\alpha^\mu\del_\mu+\beta)
\ee

Defining (in term of a generalized spin connection containing also the gauge
fields)
\begin{eqnarray}
\omega_\mu&=&\frac12 g_{\mu\nu}\left(\alpha^\nu+g^{\sigma\rho} \Gamma^\nu_{\sigma\rho}\mathbb I\right)\nonumber\\
\Omega_{\mu\nu}&=&\del_\mu\omega_\nu-\del_\nu\omega_\mu+[\omega_\mu,\omega_\nu]\nonumber\\
E&=&\beta-g^{\mu\nu}\left(\del_\mu\omega_\nu+\omega_\mu\omega_\nu-\Gamma^\rho_{\mu\nu}\omega_\rho\right)
\end{eqnarray}
then
\begin{eqnarray}
a_0&=&\frac{\Lambda^4}{16\pi^2}\int\dd x^4 \sqrt{g}
\tr\mathbb I_F\nonumber\\
a_2&=&\frac{\Lambda^2}{16\pi^2}\int\dd x^4 \sqrt{g}
\tr\left(-\frac R6+E\right)\nonumber\\
a_4&=&\frac{1}{16\pi^2}\frac{1}{360}\int\dd x^4 \sqrt{g}
\tr(-12\nabla^\mu\nabla_\mu R +5R^2-2R_{\mu\nu}R^{\mu\nu}\nonumber\\
&&+2R_{\mu\nu\sigma\rho}R^{\mu\nu\sigma\rho}-60RE+180E^2+60\nabla^\mu\nabla_\mu
E+30\Omega_{\mu\nu}\Omega^{\mu\nu}) \label{spectralcoeff}
\end{eqnarray}
$\tr$ is the trace over the inner indices of the finite algebra
$\mathcal A_F$ and  $\Omega$ and $E$  contain the gauge
degrees of freedom including the gauge stress energy tensors and the
Higgs, which is given by the inner fluctuations of $D$

Let me analyse the role of $\Lambda$. Without it, the trace in \eqref{bosonicspectralact} would diverge. Field theory cannot be valid at all scales. It is itself a theory which emerges form a yet unknown quantum gravity.
This points to a geometry in which the spectrum of operators like Dirac operator are  {truncated}, i.e.\ the eigenvalues ``saturate'' at $\Lambda$, which appears as the top scale at which one can use QFT. One may identify this scale with $\ell$, but it might be different (even lower, at the unification scale). Apart from the spectral action, truncation on a matrix basis is a common tool in noncommutative geometry~\cite{matrixreview}.

Consider the eigenvectors $\ket{n}$ of \formu{D} in increasing order of the respective eigenvalue \formu{\lambda_n}. \formu{D=\sum_0^\infty\ket n \lambda_n \bra n}. Define \formu N as the maximum value for which \formu{\lambda_n\leq\Lambda}.  This defines the truncated Dirac operator 
\be
D_\Lambda=\sum_0^N\ket n \lambda_n \bra n+\sum_N^\infty\ket n \Lambda \bra n. 
\ee
We are effectively saturating the operator at a scale \formu{\Lambda}.

Given a space with a Dirac operator one can define a distance~\cite{Connesbook} between states of the algebra of functions, in particular points are (pure) states and the distance is:
\formula{d(x,y)=\sup_{\|[D,f]\|\leq1}\left| f(x)-f(y)\right| }
It is possible possible to prove~\cite{DAndrea:2013rix}
 that using \formu{D_\Lambda}   the distance among points is infinite. In general for a bounded Dirac operator of norm \formu{\Lambda} then \formu{d(,x,y)>\Lambda^{-1}}, and to find states at finite distance one has to consider ``extended'' points, such as coherent states.

I will try to infer, form a field theory which I know works at ``low'' energy, the behaviour of it at scales which are beyond a cutoff. I am of course on dangerous grounds, I am stretching a theory beyond it natural realm of validity. I am assuming that the scale of renormalization has a physical meaning, and is not a device to regularize the infinities of the theory, and I am using an action which, while is capable to reproduce some features of the standard model, is not really equipped to quantize gravity.
Having dome the disclaimer let me proceed. I will investigate the spectral action in the limit of \emph{high momenta} using an expansion which sums up all derivatives. I will follow closely my paper with Vassilevich and Kurkov~\cite{Kuliva}.

Consider the generic fermionic action:
\formula{Z = \int [d\bar\psi][d\psi]e^{-\langle \psi | D |\psi\rangle \ } \stackrel{\mbox{\tiny formally}}{=}  \det{ D} } 
 The equality is formal because the expression is divergent, and has to be regularized, and the natural choice for us is to chose $D_\Lambda$. The considerations are nevertheless more general. One can study the renormalization flow, and note that the measure is not invariant under scale transformation, giving rise to a potential anomaly~\cite{Andrianov:2010nr, Andrianov:2011bc}.
The induced term by the flow, which takes care of the anomaly, turn out to be exactly the spectral action.

We will make the working hypothesis is that \formu{\Lambda} has a physical meaning, it is a scale indicating a phase transition, and we can try to infer some properties of the phase above \formu{\Lambda} studying the high energy limit of the action with the cutoff. We will use field theory to do this, and will find, in the end, that at high momentum Green's function, the inverse of \formu{D_\Lambda}, effectively is the identity in momentum space. More precisely, we will expand the action around high momenta, rather than low ones, as is usually done. let us see this in greater detail considering the bosonic sector.

Usually probes are bosons, hence let me consider the expansion of the spectral action in the high momentum limit.
This has been made by Barvinsky and Vilkovisky~\cite{Barvinsky:1990up} who were able to sum all derivatives (for a decreasing exponential cutoff function):
\bea
\Tr{\exp{\left(-\frac{D^2}{\Lambda^2 }\right)}} 
&\simeq&\frac{\Lambda^4}{(4\pi )^2} \int d^4x \sqrt{g}\, {\rm tr}\, \left[ 1 + \Lambda^{-2}P + \right. \nonumber\\
&&\Lambda^{-4}\left( 
R_{\mu\nu} c_1\left(-\frac{\nabla^2}{\Lambda^2}\right) R^{\mu\nu} + Rc_2\left(-\frac{\nabla^2}{\Lambda^2}\right)R  +\right.\nonumber\\
&&\left. Pc_3\left(-\frac{\nabla^2}{\Lambda^2}\right)R +Pc_4\left(-\frac{\nabla^2}{\Lambda^2}\right) P 
+ \Omega_{\mu\nu} c_5\left(-\frac{\nabla^2}{\Lambda^2}\right)\Omega^{\mu\nu}
\bigr) \right]  \nonumber\\
&&+ O(R^3,\Omega^3,E^3) 
\eea
where \formu{P= E+\tfrac 16 R} and  \formu{c_1,\ldots,c_5} are known functions, high momenta asymptotic of form factor which can be calculated and are
\bea
c_1(\xi) &\simeq& \frac{1}{6}\,{\xi}^{-1}-{\xi}^{-2} + O \left( {\xi}^
{-3} \right) \nonumber \\
c_2(\xi) &\simeq& -\frac{1}{18}\,{\xi}^{-1}+\frac{2}{9}\,{\xi}^{-2} + O \left( {\xi}^{-3}
 \right)  \nonumber \\
 c_3(\xi) &\simeq& -\frac{1}{3}\,{\xi}^{-1}+\frac{4}{3}\,{\xi}^{-2} + O
 \left( {\xi}^{-3} \right) \nonumber\\
 c_4(\xi) &\simeq& {\xi}^{-1}+2\,{\xi}^{-2} + O \left( {\xi}^{-3}
 \right)  \nonumber\\
 c_5(\xi) &\simeq& \frac{1}{2}\,{\xi}^{-1}-{\xi}^{-2} + O \left( {\xi}
^{-3} \right)
\eea

Let us first discuss the usual case (no truncation). Consider a Dirac operator containing just the relevant aspects, i.e. a bosonic fields and the fluctuations of the metric.
\formula{
\slashed{D}=i\gamma^\mu \nabla_\mu + \gamma_5 \phi= i\gamma^\mu (\del_\mu+\omega_\mu + iA_
\mu) +\gamma_5 \phi\label{sD}}
with \formu{\omega_\mu} the Levi-Civita connection and  \formu{g_{\mu\nu}=\delta_{\mu\nu}+h_{\mu\nu}}. It is now possible to perform the B-V expansion to get the expression for the high energy spectral action.
\formula{ 
S_B\simeq  
\frac {\Lambda^4}{(4\pi )^2} \int d^4x \left[ -\tfrac 32  h_{\mu\nu}h_{\mu\nu}
+8 \phi \frac 1{-\partial^2} \phi + 8 F_{\mu\nu} \frac 1{(-\partial^2)^2} F_{\mu\nu} \right] }
In order to understand the meaning of this action let me remind how we get propagation of particles and fields and correlation of points in usual QFT with action:
\formula{
S[J,\varphi] = \int d^4 x \left[\varphi(x)\left(\partial^2 + m^2 \right)\varphi(x)  - J(x)\varphi(x) \right]}
To this correspond the equation of motion
\formula{
\left(\partial^2 + m^2 \right)\varphi(y)  = J(y)}
And the Green's function \formu{G(x-y)} which ``propagates'' the source: 
\formula{
\varphi_J(x)=\int d^4y J(y) G(x-y)}
In the momentum representation we have:
\bea
\varphi(x) &=&  \frac{1}{\left(2\pi\right)^2} \int d^4 k ~e^{ikx} ~ \hat{\varphi}(k) \nonumber\\
J(x) &=& \frac{1}{\left(2\pi\right)^2} \int d^4 k ~e^{ikx} ~ \hat{J}(k)\nonumber\\
G(x-y) &=& \frac{1}{\left(2\pi\right)^2} \int d^4 k ~e^{ik(x-y)} ~ \hat{G}(k)
\eea
And the propagator is
\formula{G(k) = \frac1{\left(k^2+m^2\right)}}
The field at a point depends on the value of field in nearby points, and the points ``talk'' to each other exchanging virtual particles.

In the general case of a generic boson \formu{\varphi}, the Higgs, an intermediate vector boson or the graviton and \formu{F(\del^2)} the appropriate wave operator, a generalised Laplacian
\formula{
S[J,\phi]=\int d^4x \left(\frac12 \varphi(x) F(\del^2) \varphi(x) -J(x) \varphi(x)\right)\,,}
In this case the equation of motion is
\be
F(\del^2)\phi(x)=J(x)\label{eom}
\ee
giving
\formula{
G=\frac1{F(\del^2)}\ \ \ , \ \ G(k)=\frac1{F(-k^2)} }
and
\formula{
\varphi_J(x)=\int d^4y J(y) G(x-y)=\frac1{(2\pi)^4}\int d^4k e^{ikx}J(k)\frac1{F(-k^2)} }

The cutoff is telling us that
\formula{
\varphi_J(x)=\int d^4y J(y) G(x-y)=\frac1{(2\pi)^4}\int d^4k e^{ikx}J(k)\frac1{F(-k^2)} }
The short distance behaviour is given by the limit \formu{k\to\infty}.

Consider \formu{J(k)\neq 0} for \formu{|k^2|\in[K^2,K^2+\delta k^2]}, with \formu{K^2} very large. 
Then the Green's function becomes:
\formula{
\varphi_J(x) \xrightarrow[{{\scriptstyle K\to\infty}}]{}
\left\{\begin{array}{ll} \frac1{(2\pi)^4}\int d^k e^{ikx} J(k) k^2=(-\del^2)J(x)  & \mbox{\small for scalars and vectors}  \\
\frac1{(2\pi)^4}\int d^k e^{ikx} J(k) =J(x)  & \mbox{\small for gravitons} 
\end{array}
\right. 
}
This  corresponds to a limit of the Green's function in position space
\formula{
G(x-y)\propto\left\{\begin{array}{ll} (-\del^2)\delta\left(x-y\right)   & \mbox{for scalars and vectors}  \\
\delta\left(x-y\right)  & \mbox{for gravitons} 
\end{array}
\right. }
The correlation vanishes for noncoinciding points, heuristically, nearby points ``do not talk to each other''.
This is a limiting behaviour, we are inferring the behaviour at high energy extrapolating from low energy expanding around infinity. It is a little like a fish attempting to know what is beyond the surface of the water!
I think one has to take it as a general indication that the presence of a physical cutoff scale in momenta leads to a ``non geometric phase'' in which the concept of point ceases to have meaning, possibly described by a noncommutative geometry

Note that throughout this discussion spacetime has been left unaltered. we just imposed the cutoff and used standard techniques and interpretations. Hence we did not touch upon points in this discussion.  Points might still be there, but they are uncorrelated at very high energy. Somehow their meaning is lost, since they are not operatively correlated. In our opinion this indicates a phase transition for which the locality order parameter (position) is not present anymore.
 But it also gives us the indication that quantum gravity must be a theory in which points are not the relevant entity.
This is coherent with all other indications.
In this interpretation we are like the deep water fish trying to understand what goes on above his ceiling.
He knows that pressure decreases as he goes up. He can also infer some properties of a different states of matter by looking at bubbles which are creates near some ``high energy'' volcanic vents or when ``above'' there are storms, but he cannot naturally grasp the concept of air, or absence of water.
He will need a higher leap: quantum gravity. The hope is that this kind of considerations might help.

\subsection*{Acknowledgement}
I would like to thank Michal Heller and Michal Eckstein for organizing the meeting which gave me a chance to think with a more ``phylosophical'' perspective. The whole experience was very stimulating and I enjoyed it very much. I thank Maxim Kurkov for a critical reading of the manuscript.
I am also grateful for the support of the INFN Iniziativa Specifica GeoSymQFT and Spanish
MINECO under project MDM-2014-0369 of ICCUB (Unidad de Excelencia `Maria de Maeztu') and  {\sl COST} action \emph{QSPACE}.

\end{document}